\documentclass[fleqn,showpacs,preprint,preprintnumbers,amsmath,amssymb]{revtex4}

\usepackage{graphicx}% Include figure files
\usepackage{dcolumn}% Align table columns on decimal point
\usepackage{bm}% bold math
\begin{document}

\title{Distinctive features of ion-acoustic solitons in EPI super-dense magneto-plasmas with degenerate electrons and positrons}

\author{M. Akbari-Moghanjoughi}
\affiliation{Azarbaijan University of
Tarbiat Moallem, Faculty of Sciences,
Department of physics, 51745-406, Tabriz, Iran}

\date{\today}

\begin{abstract}
Using the extended Poincar\'{e}-Lighthill-Kuo (PLK) reductive perturbation method to study the small-amplitude ion acoustic solitary wave (IASW) dynamics (propagation and interaction), it is shown that in Thomas-Fermi magneto-plasma consisting of inertial-less degenerate electrons and positrons and isothermal ions, distinctive features emerge when the ultra-relativistic degeneracy pressure applies to electrons and positrons. Calculations show that ion-acoustic solitary waves may interact differently in such plasmas under ultra-relativistic degeneracy pressure.
\end{abstract}


\maketitle

\section{Introduction}\label{intro}

Ion acoustic wave (IAW) is one of well understood characteristics of plasma environments, investigations of which has been going on for about several decades \cite{vedenov, davidson}. Interesting nonlinear features of these quantities in different plasma environments as well as their abundant occurrence in nature has grown the researcher's attention over the past few years. One of the important nonlinear features of the kind is the ion-acoustic solitary waves (IASWs). Discovery of remarkable shape-preservation feature of these waves during their interactions by Zabusky et.al. \cite{zabusky} in 1965, made their first important applications in communication technology. Washimi et.al \cite{washimi} showed that such waves, in a weakly nonlinear regime, can be mathematically modeled by the well known Korteweg-de Vries (KdV) equations. Oikawa et.al. \cite{oikawa} have used an extended approach to consider the interaction of such waves as the superposition of two single KdV-type solitons. The method of extended reductive perturbation is one of standard methods to study the interaction of solitons in many theoretical fields including nonlinear optics, Bose-Einstein condensates, solid-states and plasma research. The technique is so-called extended Poincar\'{e}-Lighthill-Kuo (PLK) method \cite{jeffery, masa}. It has also been shown \cite{nob} that, using the method of multiple scales combined with the reductive perturbation method, not only eliminates the secularities arising in the second order correction but also the phase-factor of the lowest KdV soliton suffers a modification proportional to its amplitude.

Due to vital importance in astrophysical sciences, in the past few years the nonlinear aspects of IASW dynamics in electron-positron-ion (EPI) or more generally pair-plasma has been studied extensively using multiple scales \cite{salah, Esfand1, Esfand2, Esfand3, Tiwari1, Tiwari2, Mushtaq, akbari} and Sagdeev pseudo-potential \cite{popel, nejoh, mahmood1, mahmood2, mahmood3} techniques. EPI plasmas can be found in many astronomical objects such as active-galactic-nuclei \cite{miller}, pulsar magnetospheres \cite{michel}, neutron stars and supernovas etc. \cite{silva}. Among other examples are magnetic confinement systems and intense-laser matter interaction experiments \cite{surko1, surko2, greeves, Berezh}. On the other hand, in white dwarfs or cores of giant planets like our Jupiter, where the density of matter is much higher than that of ordinary solids and in places which can be as hot as fusion plasmas, the quantum mechanical rules govern the collective behavior of the degenerate matter \cite{Manfredi}. In such a dense plasma state, where the de Broglie thermal wavelength $\lambda_B = h/(2\pi m_e k_B T)^{1/2}$ is comparable or higher than inter-particle distances \cite{Bonitz}, there remains a question of whether the pair annihilation rate will destroy all the produced positrons. However, the Pauli exclusion principal causes the annihilation rates to be much lower in a degenerate plasma compared to ordinary EPI one, since the electron-positron interactions are limited to a narrow energy band, a process called Fermi-blocking. In other words, only the electrons and positrons with energies within $k_BT_{e,p}$ around the Fermi-energy take part in collective plasma phenomena, hence, the plasma is almost collision-less. Trivial calculations \cite{Sabry, dubinov2} (e.g. see appendix therein) confirm that in a white dwarf positrons are long lived enough to contribute in nonlinear phenomena.

On the other hand, the electron degeneracy pressure which holds the massive white-dwarfs against their gravitational pressure, softens when the electrons and positrons get relativistic leading to further gravitational collapse of the star into a black-hole \cite{chandra}. Therefore, the study of nonlinear IASW dynamics in such state of high ultra-relativistic degeneracy pressure may be of a great interest to scientists in the field. More recently, A. E. Dubinov, et. al. have developed a nonlinear theory of IASWs in ideal plasma with degenerate electrons \cite{dubinov1} and electron-positron-ion plasmas \cite{dubinov2}. The aim in current research is to investigate the effects of ultra-relativistic electron/positron degeneracy pressure on propagation and head-on collision of IASWs in a super-dense EPI magneto-plasma and address the peculiarities which may exist. This study may be helpful in understanding of nonlinear wave dynamics in super-dense magnetized stellar objects.

The article is organized in the following way. The basic normalized hydrodynamics equations are introduced in section \ref{basic}. Evolution equations along with collision parameters are derived in section \ref{shift}. Numerical analysis and discission is presented in section \ref{discussion} and final remarks are drawn in section \ref{conclusion}.

\section{Hydrodynamic description of plasma state}\label{basic}

In this study we employ the conventional hydrodynamics (HD) fluid equations to describe the collective nonlinear behavior of plasma with classical heavy ions under the Thomas-Fermi electrostatic potential of electrons and positrons and in the presence of an oblique external magnetic field. In this case, the inertial hot ions may be adequately described as an isothermal fluid. Furthermore, the degenerate electrons and positrons can be considered also as isothermal Fermi gas since the collisions are limited in this plasma due to Pauli exclusion principal as mentioned in introduction. This feature of dense plasmas also causes the pair-annihilation rate to be ignored relative to collective plasma periods in such plasmas \cite {Sabry}. It should be noted that the model used here is applicable only for the case of ions with much lower temperatures compared to characteristic Fermi temperatures of plasma, i.e., $T_{i}\ll T_{Fe,p}$. Therefore, the normalized set of HD equations may be written as
\begin{equation}\label{normal}
\begin{array}{l}
\frac{{\partial {n_i}}}{{\partial t}} + \nabla \cdot({n_i}{{\bf{V}}_i}) = 0,{{\bf{V}}_i} = {\bf{i}}{u_i} + {\bf{j}}{v_i} + {\bf{k}}{w_i}, \\
\frac{{\partial {{\bf{V}}_i}}}{{\partial t}} + ({{\bf{V}}_i}\cdot\nabla ){{\bf{V}}_i} + \nabla \varphi  + \frac{\sigma}{n_i}\nabla {n_i} + \bar \omega ({{\bf{V}}_i} \times {\bf{k}})=0, \\
{\nabla ^2}\varphi  = ({n_e} - {n_p} - {n_i}), \\
\end{array}
\end{equation}
where, the parameter $\sigma=T_i/T_{Fe}$ measures the relative ion-temperature with respect to Fermi energy of electrons, in which $T_{i}$ represents the ion-temperature and $T_{Fe}$ denotes electron Fermi-temperature. Also the new parameter ${\bar \omega }=\omega_{ci}/\omega_{pi} $ is defined as the plasma ion cyclotron frequency ($\omega_{ci}=eB_{0} /m_{i}$) normalized to the characteristic plasma frequency. In obtaining the normalized set of equations following scalings are used
\begin{equation}
\nabla\to \frac{{{c_{s}}}}{{{\omega _{pi}}}}\bar\nabla ,\hspace{3mm}t \to \frac{{\bar t}}{{{\omega _{pi}}}},\hspace{3mm}n_{i.p} \to \bar n_{i,p} n_e^{(0)},\hspace{3mm}{{{\bf{V}}_i}} \to \bar {{{\bf{V}}_i}}{c_{s}},\hspace{3mm}\varphi  \to \bar \varphi \frac{{{k_B}{T_{Fe}}}}{e}.
\end{equation}
where, ${\omega _{pi }} = \sqrt {{e^2}n_{e}^{(0)}/{\varepsilon _0}{m_i }}$ and ${c_{s }} = \sqrt {2{k_B}{T_{Fe }}/{m_i }}$ are characteristic plasma-frequency and electron quantum sound-speed, respectively, and $n_{e}^{(0)}$ denotes the electrons equilibrium density which relates to $P_{Fe}$, the electron linear Fermi-momentum through $n_e^{(0)} = \frac{{8\pi }}{{3{\hbar ^3}}}P_{Fe}^3$.

For degenerate electrons and positrons one may write the \emph{three-dimensional} normalized Thomas-Fermi density-energy relations as
\begin{equation}\label{dist}
{n_e} = {(1 + \varphi )^{\frac{3}{2}}},\hspace{3mm}{n_p} = \alpha {(1 - {\sigma _F}\varphi )^{\frac{3}{2}}},\hspace{3mm}\alpha  = \frac{{{n_{p0}}}}{{{n_{e0}}}},\hspace{3mm}\sigma_{F}=\frac{T_{Fe}}{T_{Fp}}.\\
\end{equation}
The Thomas-Fermi approximation is however valid only when the Fermi wavelength is much less than the wavelength of the ion-acoustic waves \cite{dubinov1}, which is fairly satisfied when $T_i\ll T_{Fe,p}$ in a fully degenerate Fermi gas (under the zero temperature Fermi gas assumption). Furthermore, we only consider the low-frequency ion-acoustic solitary waves (IASWs) in which $\omega\ll\omega_{ci}$ or equivalently when the ion thermal velocity is much less than the value $\omega_{ci}/k$. On the other hand, in order to contain the validity of Thomas-Fermi approximation in external magnetic field \cite{men}, $\omega_{ci}/k$ should be much less than the electron/positron Fermi-velocity.

The equilibrium charge neutrality condition is given by Poisson's relation as
\begin{equation}
\alpha  + \beta  = 1,\hspace{3mm}\beta  = \frac{{{n_{i0}}}}{{{n_{e0}}}},
\end{equation}
Now we transform the normalized plasma equations (Eqs. (\ref{normal})) to the appropriate strained coordinate defined below which admits the seperation of variables and allows successful elimination of secular terms leading to the desired evolution equations and the corresponding collision phase-shifts \cite{jeffery, washimi, masa}
\begin{equation}\label{stretch}
\begin{array}{l}
\xi  = \varepsilon (kx + ly + mz - {c_\xi }t) + {\varepsilon ^2}{P_0}(\eta ,\tau ) + {\varepsilon ^3}{P_1}(\xi ,\eta ,\tau ) +  \ldots , \\
\eta  = \varepsilon (kx + ly + mz - {c_\eta }t) + {\varepsilon ^2}{Q_0}(\xi ,\tau ) + {\varepsilon ^3}{Q_1}(\xi ,\eta ,\tau ) +  \ldots , \\
\tau  = {\varepsilon ^3}t,\hspace{3mm}{c_\xi } = c,\hspace{3mm}{c_\eta } =  - c, \\
\end{array}
\end{equation}
where, the functions $P_j$ and $Q_j$ ($j=0,1,2,...$) describe the phase records of the traveling solitary waves which will be determined later along with the wave evolution equations in the proceeding section. We note that the interacting solitons, initially far appart, travel at directions described by cosine indices $(k,l,m)$ and the direction of magnetic field with respect to this collision line is measured by $\gamma$ defined in terms of directional indices as
\begin{equation}
\begin{array}{l}
m = \cos\gamma,\\
k^2 + l^2 + m^2 = 1.\\
\end{array}
\end{equation}
Next, we expand the dependent plasma variables around their equilibrium values through smallness, positive and real $\varepsilon$ parameter, which is of the order of perturbation amplitude and is a measure of nonlinearity strength \cite{infeld}
\begin{equation}\label{Ordering}
\left[ {\begin{array}{*{20}{c}}
{{n_i}}  \\
{\begin{array}{*{20}{c}}
{{u_i}}  \\
{{v_i}}  \\
{{w_i}}  \\
\end{array}}  \\
\varphi   \\
\end{array}} \right] = \left[ {\begin{array}{*{20}{c}}
1-\alpha   \\
{\begin{array}{*{20}{c}}
0  \\
0  \\
0  \\
\end{array}}  \\
0  \\
\end{array}} \right] + {\varepsilon ^2}\left[ {\begin{array}{*{20}{c}}
{n_i^{(1)}}  \\
{\begin{array}{*{20}{c}}
0  \\
0  \\
{w_i^{(1)}}  \\
\end{array}}  \\
{{\varphi ^{(1)}}}  \\
\end{array}} \right] + {\varepsilon ^3}\left[ {\begin{array}{*{20}{c}}
{n_i^{(2)}}  \\
{\begin{array}{*{20}{c}}
{u_i^{(1)}}  \\
{v_i^{(1)}}  \\
{w_i^{(2)}}  \\
\end{array}}  \\
{{\varphi ^{(2)}}}  \\
\end{array}} \right] + {\varepsilon ^4}\left[ {\begin{array}{*{20}{c}}
{n_i^{(3)}}  \\
{\begin{array}{*{20}{c}}
{u_i^{(2)}}  \\
{v_i^{(2)}}  \\
{w_i^{(3)}}  \\
\end{array}}  \\
{{\varphi ^{(3)}}}  \\
\end{array}} \right] +  \ldots
\end{equation}
The complete set of reduced plasma equations in strained coordinate is presented in appendix A. Isolation of the lowest-orders in $\varepsilon$ yields the following relations
\begin{subequations}\label{leading}
\begin{equation}
c\left( {-\frac{\partial }{{\partial \xi }} + \frac{\partial }{{\partial \eta }}} \right)n_i^{(1)} + m(1-\alpha) \left( {\frac{\partial }{{\partial \xi }} + \frac{\partial }{{\partial \eta }}} \right)w_i^{(1)} = 0,
\end{equation}
\begin{equation}
k(1-\alpha) \left( {\frac{\partial }{{\partial \xi }} + \frac{\partial }{{\partial \eta }}} \right){\varphi ^{(1)}} + k\sigma \left( {\frac{\partial }{{\partial \xi }} + \frac{\partial }{{\partial \eta }}} \right)n_i^{(1)} - (1-\alpha) \bar \omega v_i^{(1)} = 0,
\end{equation}
\begin{equation}
l(1-\alpha) \left( {\frac{\partial }{{\partial \xi }} + \frac{\partial }{{\partial \eta }}} \right){\varphi ^{(1)}} + l\sigma \left( {\frac{\partial }{{\partial \xi }} + \frac{\partial }{{\partial \eta }}} \right)n_i^{(1)} + (1-\alpha) \bar \omega u_i^{(1)} = 0,
\end{equation}
\begin{equation}
c(1-\alpha) \left( { - \frac{\partial }{{\partial \xi }} + \frac{\partial }{{\partial \eta }}} \right)w_i^{(1)} + m(1-\alpha) \left( {\frac{\partial }{{\partial \xi }} + \frac{\partial }{{\partial \eta }}} \right){\varphi ^{(1)}} + m\sigma \left( {\frac{\partial }{{\partial \xi }} + \frac{\partial }{{\partial \eta }}} \right)n_i^{(1)} = 0,
\end{equation}
\begin{equation}
n_i^{(1)} = \frac{3}{2}(1 + \alpha {\sigma _F}){\varphi ^{(1)}},
\end{equation}
\end{subequations}
which result in the following first-order approximations
\begin{equation}\label{Firstcomp}
\begin{array}{l}
n_i^{(1)} = \frac{3}{2}(1 + \alpha {\sigma _F})\left[ {\varphi ^{(1)}(\xi ,\tau ) + \varphi ^{(1)}(\eta ,\tau )} \right], \\
u_i^{(1)} =  - \frac{{l\delta }}{{2\bar \omega (1-\alpha) }}\left[ {{\partial _\xi}\varphi ^{(1)}(\xi ,\tau ) + {\partial _\eta }\varphi ^{(1)}(\eta ,\tau )} \right], \\
v_i^{(1)} = \frac{{k\delta }}{{2\bar \omega (1-\alpha) }}\left[ {{\partial _\xi }\varphi ^{(1)}(\xi ,\tau ) + {\partial _\xi }\varphi ^{(1)}(\eta ,\tau )} \right], \\
w_i^{(1)} = \frac{{3c}}{{2m(1-\alpha) }}(1 + \alpha {\sigma _F})\left[ {\varphi ^{(1)}(\xi ,\tau ) - \varphi ^{(1)}(\eta ,\tau )} \right], \\
\delta  = 2(1-\alpha)  + 3\sigma (1 + \alpha {\sigma _F}). \\
\end{array}
\end{equation}
The solvability condition (dispersion relation) is then given by
\begin{equation}\label{dispers}
\frac{{(1-\alpha) {m^2}}}{{{c^2} - \sigma {m^2}}} = \frac{3}{2}(1 + \alpha {\sigma _F}),
\end{equation}
which results in the normalized phase-speed $c$ of waves as
\begin{equation}\label{speed}
c = \sqrt {\frac{\delta }{{3(1 + \alpha {\sigma _F})}}} \cos \gamma.
\end{equation}
By taking into account the next higher-order in $\varepsilon$, we get the second-order approximations. Therefore, The second-order coupled differential equations are of the following forms
\textbf{}
\begin{subequations}\label{second}
\begin{equation}
\begin{array}{l}
c(1-\alpha) \left( { - \frac{\partial }{{\partial \xi }} + \frac{\partial }{{\partial \eta }}} \right)n_i^{(2)} + (1-\alpha) k\left( {\frac{\partial }{{\partial \xi }} + \frac{\partial }{{\partial \eta }}} \right)u_i^{(1)} +  \\
(1-\alpha) l\left( {\frac{\partial }{{\partial \xi }} + \frac{\partial }{{\partial \eta }}} \right)v_i^{(1)} + m(1-\alpha) \left( {\frac{\partial }{{\partial \xi }} + \frac{\partial }{{\partial \eta }}} \right)w_i^{(2)} = 0, \\
\end{array}
\end{equation}
\begin{equation}
\begin{array}{l}
c(1-\alpha) \left( { - \frac{\partial }{{\partial \xi }} + \frac{\partial }{{\partial \eta }}} \right)u_i^{(1)} + k(1-\alpha) \left( {\frac{\partial }{{\partial \xi }} + \frac{\partial }{{\partial \eta }}} \right){\varphi ^{(2)}} +  \\
k\sigma \left( {\frac{\partial }{{\partial \xi }} + \frac{\partial }{{\partial \eta }}} \right)n_i^{(2)} - (1-\alpha) \bar \omega v_i^{(2)} = 0, \\
\end{array}
\end{equation}
\begin{equation}
\begin{array}{l}
c(1-\alpha) \left( { - \frac{\partial }{{\partial \xi }} + \frac{\partial }{{\partial \eta }}} \right)v_i^{(1)} + l(1-\alpha) \left( {\frac{\partial }{{\partial \xi }} + \frac{\partial }{{\partial \eta }}} \right){\varphi ^{(2)}} +  \\
l\sigma \left( {\frac{\partial }{{\partial \xi }} + \frac{\partial }{{\partial \eta }}} \right)n_i^{(2)} + (1-\alpha) \bar \omega u_i^{(2)} = 0, \\
\end{array}
\end{equation}
\begin{equation}
\begin{array}{l}
c(1-\alpha) \left( { - \frac{\partial }{{\partial \xi }} + \frac{\partial }{{\partial \eta }}} \right)w_i^{(2)} + m(1-\alpha) \left( {\frac{\partial }{{\partial \xi }} + \frac{\partial }{{\partial \eta }}} \right){\varphi ^{(2)}} + \\
m\sigma \left( {\frac{\partial }{{\partial \xi }} + \frac{\partial }{{\partial \eta }}} \right)n_i^{(2)} = 0,
\end{array}
\end{equation}
\begin{equation}
n_i^{(2)} = \frac{3}{2}(1 + \alpha {\sigma _F}){\varphi ^{(2)}},
\end{equation}
\end{subequations}
which, consequently, yield the following second-order components
\begin{equation}\label{uv2}
\begin{array}{l}
n_i^{(2)} = \frac{3}{2}(1 + \alpha {\sigma _F})\left[ {\varphi ^{(2)}(\xi ,\tau ) + \varphi ^{(2)}(\eta ,\tau )} \right], \\
u_i^{(2)} = \frac{{ck\delta }}{{2{{\bar \omega }^2}(1-\alpha) }}\left[ {{\partial _\xi }\varphi ^{(2)}(\xi ,\tau ) - {\partial _\eta }\varphi ^{(2)}(\eta ,\tau )} \right] - \frac{{l\delta }}{{2\bar \omega (1-\alpha) }}\left[ {{\partial _{\xi \xi }}\varphi ^{(1)}(\xi ,\tau ) + {\partial _{\eta \eta }}\varphi ^{(1)}(\eta ,\tau )} \right], \\
v_i^{(2)} = \frac{{cl\delta }}{{2{{\bar \omega }^2}(1-\alpha) }}\left[ {{\partial _\xi }\varphi ^{(2)}(\xi ,\tau ) - {\partial _\eta }\varphi ^{(2)}(\eta ,\tau )} \right] + \frac{{k\delta }}{{2\bar \omega (1-\alpha) }}\left[ {{\partial _{\xi \xi }}\varphi ^{(1)}(\xi ,\tau ) + {\partial _{\eta \eta }}\varphi ^{(1)}(\eta ,\tau )} \right], \\
w_i^{(2)} = \frac{{3c}}{{2m(1-\alpha) }}(1 + \alpha {\sigma _F})\left[ {\varphi ^{(2)}(\xi ,\tau ) - \varphi ^{(2)}(\eta ,\tau )} \right]. \\
\end{array}
\end{equation}
where, ${\varphi ^{(1)}}(\xi,\tau)$ and ${\varphi ^{(1)}}(\eta,\tau)$ describe the first-order amplitude evolution and ${\varphi ^{(2)}}(\xi,\tau)$ and ${\varphi ^{(2)}}(\eta,\tau)$ describe the second-order amplitude evolution components of two distinct solitary excitations in the oblique directions ${\eta_ \bot }$ and ${\xi_ \bot }$ (${\eta_ \bot=-\xi_ \bot }$), respectively. In forthcoming algebra we will use the notations ${\varphi_\xi ^{(1)}}$ and ${\varphi_\eta ^{(1)}}$ instead of ${\varphi ^{(1)}}(\xi,\tau)$ and ${\varphi ^{(1)}}(\eta,\tau)$ for simplicity.

\section{Solitary Wave Dynamics Description}\label{shift}

In order to obtain the third-order approximation for density component we consider the next higher terms in $\varepsilon$. Again by solving the coupled differential equations in this approximation level and by making use of compatibility relation (Eq. \ref{dispers}) and the first- and second-order plasma approximations, we obtain
\begin{equation}\label{n3}
\begin{array}{l}
n_i ^{(3)} ={K}{N}\left[ {\frac{{\partial \varphi _\eta^{(1)}}}{{\partial \tau}} + A\varphi _\eta^{(1)}\frac{{\partial \varphi _\eta^{(1)}}}{{\partial \eta}} - B\frac{{{\partial ^3}\varphi _\eta^{(1)}}}{{\partial {\eta^3}}}} \right]\xi -
{K}{N}\left[ {\frac{{\partial \varphi _\xi^{(1)}}}{{\partial \tau}} - A\varphi _\xi^{(1)}\frac{{\partial \varphi _\xi^{(1)}}}{{\partial \xi}} + B\frac{{{\partial ^3}\varphi _\xi^{(1)}}}{{\partial {\xi^3}}}} \right]\eta +  \\
{K{E_2}}\left[ {{P_0}(\eta,\tau) - \frac{E_1}{E_2}\int {\varphi _\eta^{(1)}d\eta} } \right]\frac{{\partial \varphi _\xi^{(1)}}}{{\partial \xi}} -
{K{E_2}}\left[ {{Q_0}(\xi,\tau) - \frac{E_1}{E_2}\int {\varphi _\xi^{(1)}d\xi} } \right]\frac{{\partial \varphi _\eta^{(1)}}}{{\partial \eta}} +  \\
{K}{N}\left[ {\int {\frac{{\partial \varphi _\xi^{(1)}}}{{\partial \tau}}d\xi}  - \int {\frac{{\partial \varphi _\eta^{(1)}}}{{\partial \tau}}d\eta}}\right] - {CK}\left[ {{{(\varphi _\xi^{(1)})}^2} - {{(\varphi _\eta^{(1)})}^2}} \right] + DK\left[ {\frac{{{\partial ^2}\varphi _\xi^{(1)}}}{{\partial {\xi^2}}} - \frac{{{\partial ^2}\varphi _\eta^{(1)}}}{{\partial {\eta^2}}}} \right] + \\
F(\xi,\tau) + G(\eta,\tau), \\
\end{array}
\end{equation}
where, the functions $F(\xi,\tau)$ and $G(\eta,\tau)$ are the homogenous solutions of differential equations in this order. The other unknown coefficients in Eq. (\ref{n3}) are as follows
\begin{subequations}\label{coeffs}
\begin{equation}
K = \frac{1-\alpha }{{2\delta }},
\end{equation}
\begin{equation}
A = \frac{{9\sigma {{(1 + \alpha {\sigma _F})}^3} + 2{(1-\alpha) ^2}(1 - \alpha {\sigma _F}^2)}}{{4(1-\alpha) {{(1 + \alpha {\sigma _F})}^2}}}\sqrt {\frac{{1 + \alpha {\sigma _F}}}{{3\delta }}} \cos\gamma,
\end{equation}
\begin{equation}
B = \frac{{4(1-\alpha) {\bar\omega ^2} + {\delta ^2}{{\sin }^2}\gamma }}{{6\bar\omega^2 {{(1 + \alpha {\sigma _F})}^2}}}\sqrt {\frac{{1 + \alpha {\sigma _F}}}{{3\delta }}} \cos \gamma,
\end{equation}
\begin{equation}
N = \frac{{3{{(1 + \alpha {\sigma _F})}^2}}}{{2(1-\alpha) \cos \gamma }}\sqrt {\frac{{3\delta }}{{1 + \alpha {\sigma _F}}}},
\end{equation}
\begin{equation}
E_1 = \frac{{3\left[ {9\sigma {{(1 + \alpha {\sigma _F})}^3} - 2{(1-\alpha) ^2}(1 - \alpha {\sigma _F}^2)} \right]}}{{8{(1-\alpha) ^2}}},
\end{equation}
\begin{equation}
{E_2} = \frac{{3\delta (1 + \alpha {\sigma _F})}}{1-\alpha },
\end{equation}
\begin{equation}
C = \frac{{3\left[ {9\sigma {{(1 + \alpha {\sigma _F})}^3} + 2{(1-\alpha) ^2}(1 - \alpha {\sigma _F}^2)} \right]}}{{16{(1-\alpha) ^2}}},
\end{equation}
\begin{equation}
D = \frac{{{\delta ^2}\sin^2\gamma }}{{4(1-\alpha) {\bar\omega ^2}}} - 1,
\end{equation}
\end{subequations}
The two first terms in Eqs. (\ref{n3}) are secular, because they diverge as $\xi\rightarrow\pm\infty$ and $\eta\rightarrow\pm\infty$, and they must vanish. Consequently, two distinct KdV evolution equations one for each traveling IA wave are obtained. On the other hand, the next two terms in Eq. (\ref{n3}) may cause spurious resonances \cite{nob, masa} at the next higher-order and their coefficients must also vanish. This condition determines the phase functions introduced in Eqs. (\ref{stretch}). Therefore, the full determination of dynamics of the interacting solitary waves is given by the following coupled differential equations for each wave
\begin{equation}\label{kdv1}
\frac{{\partial \varphi _\xi^{(1)}}}{{\partial \tau}} + A\varphi _\xi^{(1)}\frac{{\partial \varphi _\xi^{(1)}}}{{\partial \xi}} - B\frac{{{\partial ^3}\varphi _\xi^{(1)}}}{{\partial {\xi^3}}} = 0,
\end{equation}
\begin{equation}\label{P0}
{P_0}(\eta,\tau) = \frac{{{E_1}}}{{{E_2}}}\int {\varphi _\eta^{(1)}d\eta},
\end{equation}
\begin{equation}\label{kdv2}
\frac{{\partial \varphi _\eta^{(1)}}}{{\partial \tau}} - A\varphi _\eta^{(1)}\frac{{\partial \varphi _\eta^{(1)}}}{{\partial \eta}} + B\frac{{{\partial ^3}\varphi _\eta^{(1)}}}{{\partial {\eta^3}}} = 0,
\end{equation}
\begin{equation}\label{Q0}
{Q_0}(\xi,\tau) = \frac{{{E_1}}}{{{E_2}}}\int {\varphi _\xi^{(1)}d\xi},
\end{equation}
In order to obtain single-soliton solutions for Eqs. (\ref{kdv1}) and (\ref{kdv2}) which have multi-soliton solutions, we require that the perturbed potential components and their derivatives vanish at infinity, i.e.
\begin{equation}\label{boundary}
\begin{array}{l}
\mathop {\lim }\limits_{\zeta \to \pm\infty } \{\varphi _\zeta^{(1)},\frac{\partial \varphi _\zeta^{(1)}}{\partial \zeta },\frac{\partial ^2\varphi _\zeta^{(1)}}{\partial \zeta
^2}\}=0,\hspace{3mm} \zeta={\xi,\eta}.
\end{array}
\end{equation}
Consequently, we obtain
\begin{equation}\label{phi-x}
\begin{array}{l}
{\varphi _\xi ^{(1)} = \frac{{{\varphi _{\xi 0}}}}{{\cosh^2(\frac{{\xi  - {u_{\xi 0}}\tau }}{{{\Delta _\xi }}})}},}  \\
{{\varphi _{\xi 0}} = \frac{{3{u_{\xi 0}}}}{{{A}}},{\Delta _\xi } = {{(\frac{{4{B}}}{{{u_{\xi 0}}}})}^{\frac{1}{2}}},}  \\
\end{array}
\end{equation}
\begin{equation}\label{phi-y}
\begin{array}{l}
\varphi _\eta^{(1)} = \frac{{{\varphi _{\eta0}}}}{{\cosh^2(\frac{{\eta + {u_{\eta0}}\tau}}{{{\Delta _\eta}}})}},\\
{\varphi _{\eta0}} = \frac{{3{u_{\eta0}}}}{A},\hspace{3mm} {\Delta _\eta} = {(\frac{{4B}}{{{u_{\eta0}}}})^{\frac{1}{2}}}.
\end{array}
\end{equation}
where, $\varphi _{0}$ and $\Delta$ represent the soliton amplitude and width, respectively, and $u_{0}$ is the Mach-speed.

Inspection of Eqs. (\ref{coeffs}), reveals that the KdV coefficients $A$ and $B$ in Eqs. (\ref{kdv1}) and (\ref{kdv2}) change sign at $\gamma=\pi/2$, hence this critical value determines weather the solitons are compressive or rarefactive. It is further remarked that the soliton amplitude is independent of the strength of magnetic field, while it strictly depends on applied field direction, $\cos\gamma$. However, the soliton width depends on both magnitude and direction of the external magnetic field.

The collision phase-shifts of solitary IA excitations are obtained using Eqs. (\ref{P0}) and (\ref{Q0}) and the KdV solutions (Eqs. (\ref{phi-x}) and (\ref{phi-y})) as
\begin{equation}\label{phase-x}
\begin{array}{l}
{P_0}(\eta,\tau) = \frac{E_1}{E_2} \varphi _{\eta0} \Delta _\eta\tanh(\frac{{\eta - {u_{\eta0}}\tau}}{{{\Delta _\eta}}}),
\end{array}
\end{equation}
\begin{equation}\label{phase-y}
\begin{array}{l}
{Q_0}(\xi,\tau) = \frac{E'_1}{E'_2} \varphi _{\xi0} \Delta _\xi\tanh(\frac{{\xi + {u_{\xi0}}\tau}}{{{\Delta _\xi}}}).
\end{array}
\end{equation}
Hence, up to order $O(\varepsilon^2)$, we have
\begin{equation}\label{trajectory}
\begin{array}{l}
\xi = \varepsilon ({k}x + {l}y + {m}z + {c}t) - \varepsilon^2\frac{{{E_1}}}{{{E_2}}}{\varphi _{\eta0}}{\Delta _\eta}\tanh (\frac{{\eta - {u_{\eta0}}\tau}}{{{\Delta _\eta}}})+O(\varepsilon^3), \\
\eta = \varepsilon ({k}x + {l}y + {m}z - {c}t)] - \varepsilon^2\frac{{{{E'}_1}}}{{{{E'}_2}}}{\varphi _{\xi0}}{\Delta _\xi}\tanh (\frac{{\xi + {u_{\xi0}}\tau}}{{{\Delta _\xi}}})+O(\varepsilon^3), \\
\end{array}
\end{equation}

We may obtain the overall phase-shifts by comparing the phases of each wave long before and after the collision in the following manner
\begin{equation}
\begin{array}{l}
\Delta {P_0} = P_{post-collision}-P_{past-collision}=\\ \mathop {\lim }\limits_{\xi=0,\eta \to  + \infty } [\varepsilon ({k}x + {l}y + {m}z + {c}t)]-
\mathop {\lim }\limits_{\xi=0,\eta \to  - \infty } [\varepsilon ({k}x + {l}y + {m}z + {c}t)] , \\
\Delta {Q_0} = Q_{post-collision}-Q_{past-collision}=\\ \mathop {\lim }\limits_{\eta=0,\xi \to  + \infty } [\varepsilon ({k}x + {l}y + {m}z - {c}t)]-
\mathop {\lim }\limits_{\eta=0,\xi \to  - \infty } [\varepsilon ({k}x + {l}y + {m}z - {c}t)] , \\
\end{array}
\end{equation}
The phase quantities, $\Delta {P_0}$ and $\Delta {Q_0}$ present the overall phase-shifts of solitary structures $"S1"$ and $"S2"$ in a head-on collision. By making use of Eqs. (\ref{phase-x}), (\ref{phase-y}) and (\ref{stretch}) we obtain the following expressions
\begin{equation}\label{shifts}
\begin{array}{l}
\Delta {P_0} = - {\varepsilon ^2}\left[\frac{{2{(1-\alpha) ^2}(1 - \alpha {\sigma _F}^2) - 9\sigma {{(1 + \alpha {\sigma _F})}^3}}}{{4\delta(1-\alpha)  (1 + \alpha {\sigma _F})}}\right]{\varphi _{\eta 0}}{\Delta _\eta }, \\
\Delta {Q_0} = {\varepsilon ^2}\left[\frac{{2{(1-\alpha) ^2}(1 - \alpha {\sigma _F}^2) - 9\sigma {{(1 + \alpha {\sigma _F})}^3}}}{{4\delta(1-\alpha)  (1 + \alpha {\sigma _F})}}\right]{\varphi _{\xi 0}}{\Delta _\xi }. \\
\end{array}
\end{equation}

\section{Analysis and Discussions}\label{discussion}

In this section the numerical analysis is presented for two cases of plasma under degeneracy pressure, namely, when electron/positrons are normally degenerate and when they possess ultra-relativistic degeneracy. The normal degeneracy may occur for normal solid state densities ($n_e\ll 10^{35}m^{-3}$) such as that for ordinary metallic electrons, while ultra-relativistic degeneracy may be encountered in super-dense stellar objects or giant planet-interiors like Jupiter in which the densities may be much higher ($n_e\sim10^{35}m^{-3}$). From the standard definitions, it is noted that, in a three-dimensional \emph{non-relativistic} Fermi-gas for normally degenerate electrons and positrons we have $E_{Fj}=\frac{\hbar^2 k_{Fj}^2}{2m_j}$ ($j=e,p$) or $E_{Fj}\propto n_{j0}^{2/3}$, which follows (in our model) that $\sigma_F=\alpha^{-2/3}$. On the other hand, in a three-dimensional \emph{ultra-relativistic} Fermi-gas, we have $E_{Fj}=c\hbar k_{Fj}$ or $E_{Fj}\propto n_{j0}^{1/3}$ \cite{chandra}, which in our analysis means that $\sigma_F=\alpha^{-1/3}$.

Figure (1) shows the variations of soliton amplitude with respect to the magnetic field direction, $\gamma$. The variations are quite similar to those of non-degenerate EPI plasma \cite{Esfand2}. It is remarked from Fig. 1(a) and 1(b) that for plasma with both normal and relativistic degeneracy pressure below the critical value of $\gamma_{cr}=\pi/2$ the IA solitons can be either rarefactive or compressive. However, changing $\alpha$ from $0.2$ to $0.5$ changes the IASW profile from compressive to rarefactive for normal degeneracy case. This means that a critical value of fractional positron-to-electron density, $\alpha_{cr}$, also exists which defines the type of IA solitons for normal degeneracy case. For the case of normal degeneracy, three different values of $\alpha_{cr}$ is shown in Fig. 1(c). \emph{However, Fig 1(d) reveals that for the plasma with ultra-relativistic degeneracy, contrary to the normal degeneracy case, the shape of IA solitary waves is defined only by the value of $\gamma_{cr}$, and no such critical fractional positron-to-electron density, $\alpha_{cr}$ exists for this case, however, there exists a maximum amplitude at some $\alpha$-values, instead}.

Figure 2 depicts the dependencies of soliton width on different plasma parameters. It is observed from Figs. 2(a, b, c, d) that, maximum values of the soliton width occurs for some $\gamma$-values in all plots. These values of the maximum soliton width for the range of $0\geq\gamma>\pi/2$ are given by the following analytical expression
\begin{equation}\label{gamma}
{\gamma _m} = \arccos{\left[ {\frac{1}{3} + \frac{{4(1-\alpha) {\bar\omega ^2}}}{{3{\delta ^2}}}} \right]^{\frac{1}{2}}},
\end{equation}
and at the cold-ion limit we obtain
\begin{equation}\label{gamma}
{\gamma _m} = \arccos {\left[ {\frac{1}{3} + \frac{{{{\bar \omega }^2}}}{{3(1-\alpha) }}} \right]^{\frac{1}{2}}},
\end{equation}
where the maximum value of this quantity itself is ${\gamma _{mm}} \approx {54.73^ \circ }$. Moreover, it is observed that, the maximum value of IA soliton width decreases with increase of magnetic field strength, $\bar\omega$ (Fig. 2(a)) and fractional positron density, $\alpha$ (Fig. 2(b)), however, it increases with increase of fractional ion-to-electron Fermi temperature, $\sigma$ (Fig. 2(c)), when other plasma parameters are fixed. Figure 2(d) indicates that the values of soliton width for identical other plasma parameters is always higher for ultra-relativistic degeneracy case (solid-line) compared to that of normal degeneracy case (dashed-line).

Regarding the collision phase-shift variation with respect to magnetic field strength, it is revealed from Fig. 3 that, for normal degeneracy case (dashed-line) the phase shift decreases monotonically with increase of field strength, while for ultra-relativistic degeneracy case (solid-line) the phase shift suddenly drops to a small value at first and then remains almost unchanged with increase of magnetic field strength. A typical variation of collision phase-shift with magnetic field direction (Fig. 3(b)) shows that the value of phase-shift increases as the critical value of $\gamma=\pi/2$ is approached when the other parameters are fixed. \emph{On the other hand, Figs. 3(c,d) reveal another distinctive difference between the variation of the collision phase-shift of normal and ultra-relativistic degeneracy cases, so that, by approaching the critical $\sigma$- or $\alpha$-values the collision phase-shift vanishes for ultra-relativistic degeneracy case (solid-lines), while, it diverges for normal degeneracy case (dashed-lines).}

Finally, Fig. 4 depicts the areas in $\alpha$-$\sigma$ plane the regions (in-grey) in which the soliton is dark-type for the cases of $\pi/2>\gamma$ and where the head-on collision phases-shift is positive in the range $\pi/2>\gamma\geq0$ for normal and ultra relativistic degeneracy cases. It is remarked from Fig. 4(a) that for normal degeneracy case the soliton is compressive (rarefactive ) below (above) a critical $\sigma_{cr}$-value (or equivalently $\alpha_{cr}$-value) which is defined by the simple relation below
\begin{equation}\label{sigma}
{\sigma_{cr}} = \frac{{2{(1-\alpha) ^2}(1 - \alpha {\sigma _F}^2)}}{{9{{(1 + \alpha {\sigma _F})}^3}}}.
\end{equation}
\emph{It is observed from Fig. 4(b) that for ultra-relativistic degeneracy case, regardless of the value of $\sigma$ or $\alpha$, the IA solitons are always bright (dark) for $\pi/2>\gamma$ ($\gamma>\pi/2$)}. On the other hand, Figs. 4(c,d) show the region (in grey) for which the collision phase-shift is positive for a given value of $\alpha$ and $\sigma$. Therefore, it is concluded that, the value of collision phase shift changes the sign at a critical $\sigma$ ($\alpha$) value. The values of $\sigma_{cr}$ for which the collision phase-shift changes the sign is also given by Eq. (\ref{sigma}). \emph{It is also important to note (from Eq. (\ref{sigma})) that for normal relativistic degeneracy case the collision phase-shift at $\alpha=0$ (particular case of the degenerate electron-ion plasma) is always positive, while, for the case of ultra-relativistic degenerate electron-ion plasma the phase-shift is positive only in the range $0\geq\sigma\geq0.\bar2$ and negative otherwise.}

\section{Conclusions}\label{conclusion}

The extended Poincar\'{e}-Lighthill-Kuo (PLK) reductive perturbation method was used to investigate the propagation and head-on collision of ion-acoustic (IA) solitary waves in Thomas-Fermi electron-positron-ion magneto-plasma. It was shown that the relativistic degeneracy pressure of electrons and positrons in such plasma has significant effects on the propagation as well as head-on collisions of IASWs in magneto-dense degenerate plasmas. Calculations reveal that IASWs behave much different in such plasmas under non-relativistic and ultra-relativistic degeneracy pressure.

\textbf{Acknowledgement: I would like to thank Dr. A. Esfandyari-Kalejahi from whom I have learned much about the classical hydrodynamics. Great thanks are also devoted to the referee whose detailed comments has led the article to its current state.}

\appendix

\section{Normalized plasma equations in strained coordinate}
\begin{equation}\label{strain1}
\begin{array}{l}
{\varepsilon ^2}\frac{{\partial {n_i}}}{{\partial \tau }} - c\frac{{\partial {n_i}}}{{\partial \xi }} - {\varepsilon ^2}c\frac{{\partial {Q_0}}}{{\partial \xi }}\frac{{\partial {n_i}}}{{\partial \eta }} + c\frac{{\partial {n_i}}}{{\partial \eta }} + {\varepsilon ^2}c\frac{{\partial {P_0}}}{{\partial \eta }}\frac{{\partial {n_i}}}{{\partial \xi }} + k\frac{{\partial {n_i}{u_i}}}{{\partial \xi }} +\\ {\varepsilon ^2}k\frac{{\partial {Q_0}}}{{\partial \xi }}\frac{{\partial {n_i}{u_i}}}{{\partial \eta }} + k\frac{{\partial {n_i}{u_i}}}{{\partial \eta }} + {\varepsilon ^2}k\frac{{\partial {P_0}}}{{\partial \eta }}\frac{{\partial {n_i}{u_i}}}{{\partial \xi }} + l\frac{{\partial {n_i}{v_i}}}{{\partial \xi }} + {\varepsilon ^2}l\frac{{\partial {Q_0}}}{{\partial \xi }}\frac{{\partial {n_i}{v_i}}}{{\partial \eta }} +\\ l\frac{{\partial {n_i}{v_i}}}{{\partial \eta }} + {\varepsilon ^2}l\frac{{\partial {P_0}}}{{\partial \eta }}\frac{{\partial {n_i}{v_i}}}{{\partial \xi }} + m\frac{{\partial {n_i}{w_i}}}{{\partial \xi }} + {\varepsilon ^2}m\frac{{\partial {Q_0}}}{{\partial \xi }}\frac{{\partial {n_i}{w_i}}}{{\partial \eta }} + m\frac{{\partial {n_i}{w_i}}}{{\partial \eta }} +\\ {\varepsilon ^2}m\frac{{\partial {P_0}}}{{\partial \eta }}\frac{{\partial {n_i}{w_i}}}{{\partial \xi }} + \ldots  = 0, \\
\end{array}
\end{equation}
\begin{equation}\label{strain2}
\begin{array}{l}
{\varepsilon ^2}\frac{{\partial {u_i}}}{{\partial \tau }} - c\frac{{\partial {u_i}}}{{\partial \xi }} - {\varepsilon ^2}c\frac{{\partial {Q_0}}}{{\partial \xi }}\frac{{\partial {u_i}}}{{\partial \eta }} + c\frac{{\partial {u_i}}}{{\partial \eta }} + {\varepsilon ^2}c\frac{{\partial {P_0}}}{{\partial \eta }}\frac{{\partial {u_i}}}{{\partial \xi }} + k{u_i}\frac{{\partial {u_i}}}{{\partial \xi }} +  \\
{\varepsilon ^2}k{u_i}\frac{{\partial {Q_0}}}{{\partial \xi }}\frac{{\partial {u_i}}}{{\partial \eta }} + k{u_i}\frac{{\partial {u_i}}}{{\partial \eta }} + {\varepsilon ^2}k{u_i}\frac{{\partial {P_0}}}{{\partial \eta }}\frac{{\partial {u_i}}}{{\partial \xi }} + l{v_i}\frac{{\partial {u_i}}}{{\partial \xi }} + {\varepsilon ^2}l{v_i}\frac{{\partial {Q_0}}}{{\partial \xi }}\frac{{\partial {u_i}}}{{\partial \eta }} +  \\
l{v_i}\frac{{\partial {u_i}}}{{\partial \eta }} + {\varepsilon ^2}l{v_i}\frac{{\partial {P_0}}}{{\partial \eta }}\frac{{\partial {u_i}}}{{\partial \xi }} + m{w_i}\frac{{\partial {u_i}}}{{\partial \xi }} + {\varepsilon ^2}m{w_i}\frac{{\partial {Q_0}}}{{\partial \xi }}\frac{{\partial {u_i}}}{{\partial \eta }} + m{w_i}\frac{{\partial {u_i}}}{{\partial \eta }} +  \\
{\varepsilon ^2}m{w_i}\frac{{\partial {P_0}}}{{\partial \eta }}\frac{{\partial {u_i}}}{{\partial \xi }} + k\frac{{\partial \varphi }}{{\partial \xi }} + {\varepsilon ^2}k\frac{{\partial {Q_0}}}{{\partial \xi }}\frac{{\partial \varphi }}{{\partial \eta }} + k\frac{{\partial \varphi }}{{\partial \eta }} + {\varepsilon ^2}k\frac{{\partial {P_0}}}{{\partial \eta }}\frac{{\partial \varphi }}{{\partial \xi }} +  \\
k\frac{\sigma}{n_i}\frac{{\partial {n_i}}}{{\partial \xi }} + {\varepsilon ^2}k\frac{\sigma}{n_i}\frac{{\partial {Q_0}}}{{\partial \xi }}\frac{{\partial {n_i}}}{{\partial \eta }} + k\frac{\sigma}{n_i}\frac{{\partial {n_i}}}{{\partial \eta }} + {\varepsilon ^2}k\frac{\sigma}{n_i}\frac{{\partial {P_0}}}{{\partial \eta }}\frac{{\partial {n_i}}}{{\partial \xi }} -\\ \frac{{\bar \omega {v_i}}}{\varepsilon } + \ldots  = 0, \\
\end{array}
\end{equation}
\begin{equation}\label{strain3}
\begin{array}{l}
{\varepsilon ^2}\frac{{\partial {v_i}}}{{\partial \tau }} - c\frac{{\partial {v_i}}}{{\partial \xi }} - {\varepsilon ^2}c\frac{{\partial {Q_0}}}{{\partial \xi }}\frac{{\partial {v_i}}}{{\partial \eta }} + c\frac{{\partial {v_i}}}{{\partial \eta }} + {\varepsilon ^2}c\frac{{\partial {P_0}}}{{\partial \eta }}\frac{{\partial {v_i}}}{{\partial \xi }} + k{u_i}\frac{{\partial {v_i}}}{{\partial \xi }} +  \\
{\varepsilon ^2}k{u_i}\frac{{\partial {Q_0}}}{{\partial \xi }}\frac{{\partial {v_i}}}{{\partial \eta }} + k{u_i}\frac{{\partial {v_i}}}{{\partial \eta }} + {\varepsilon ^2}k{u_i}\frac{{\partial {P_0}}}{{\partial \eta }}\frac{{\partial {v_i}}}{{\partial \xi }} + l{v_i}\frac{{\partial {v_i}}}{{\partial \xi }} + {\varepsilon ^2}l{v_i}\frac{{\partial {Q_0}}}{{\partial \xi }}\frac{{\partial {v_i}}}{{\partial \eta }} +  \\
l{v_i}\frac{{\partial {v_i}}}{{\partial \eta }} + {\varepsilon ^2}l{v_i}\frac{{\partial {P_0}}}{{\partial \eta }}\frac{{\partial {v_i}}}{{\partial \xi }} + m{w_i}\frac{{\partial {v_i}}}{{\partial \xi }} + {\varepsilon ^2}m{w_i}\frac{{\partial {Q_0}}}{{\partial \xi }}\frac{{\partial {v_i}}}{{\partial \eta }} + m{w_i}\frac{{\partial {v_i}}}{{\partial \eta }} +  \\
{\varepsilon ^2}m{w_i}\frac{{\partial {P_0}}}{{\partial \eta }}\frac{{\partial {v_i}}}{{\partial \xi }} + l\frac{{\partial \varphi }}{{\partial \xi }} + {\varepsilon ^2}l\frac{{\partial {Q_0}}}{{\partial \xi }}\frac{{\partial \varphi }}{{\partial \eta }} + l\frac{{\partial \varphi }}{{\partial \eta }} + {\varepsilon ^2}l\frac{{\partial {P_0}}}{{\partial \eta }}\frac{{\partial \varphi }}{{\partial \xi }} +  \\
l\frac{\sigma}{n_i}\frac{{\partial {n_i}}}{{\partial \xi }} + {\varepsilon ^2}l\frac{\sigma}{n_i}\frac{{\partial {Q_0}}}{{\partial \xi }}\frac{{\partial {n_i}}}{{\partial \eta }} + l\frac{\sigma}{n_i}\frac{{\partial {n_i}}}{{\partial \eta }} + {\varepsilon ^2}l\frac{\sigma}{n_i}\frac{{\partial {P_0}}}{{\partial \eta }}\frac{{\partial {n_i}}}{{\partial \xi }} +\\ \frac{{\bar \omega {u_i}}}{\varepsilon } + \ldots  = 0, \\
\end{array}
\end{equation}
\begin{equation}\label{strain4}
\begin{array}{l}
{\varepsilon ^2}\frac{{\partial {w_i}}}{{\partial \tau }} - c\frac{{\partial {w_i}}}{{\partial \xi }} - {\varepsilon ^2}c\frac{{\partial {Q_0}}}{{\partial \xi }}\frac{{\partial {w_i}}}{{\partial \eta }} + c\frac{{\partial {w_i}}}{{\partial \eta }} + {\varepsilon ^2}c\frac{{\partial {P_0}}}{{\partial \eta }}\frac{{\partial {w_i}}}{{\partial \xi }} + k{u_i}\frac{{\partial {w_i}}}{{\partial \xi }} +  \\
{\varepsilon ^2}k{u_i}\frac{{\partial {Q_0}}}{{\partial \xi }}\frac{{\partial {w_i}}}{{\partial \eta }} + k{u_i}\frac{{\partial {w_i}}}{{\partial \eta }} + {\varepsilon ^2}k{u_i}\frac{{\partial {P_0}}}{{\partial \eta }}\frac{{\partial {w_i}}}{{\partial \xi }} + l{v_i}\frac{{\partial {w_i}}}{{\partial \xi }} + {\varepsilon ^2}l{v_i}\frac{{\partial {Q_0}}}{{\partial \xi }}\frac{{\partial {w_i}}}{{\partial \eta }} +  \\
l{v_i}\frac{{\partial {w_i}}}{{\partial \eta }} + {\varepsilon ^2}l{v_i}\frac{{\partial {P_0}}}{{\partial \eta }}\frac{{\partial {w_i}}}{{\partial \xi }} + m{w_i}\frac{{\partial {w_i}}}{{\partial \xi }} + {\varepsilon ^2}m{w_i}\frac{{\partial {Q_0}}}{{\partial \xi }}\frac{{\partial {w_i}}}{{\partial \eta }} + m{w_i}\frac{{\partial {w_i}}}{{\partial \eta }} +  \\
{\varepsilon ^2}m{w_i}\frac{{\partial {P_0}}}{{\partial \eta }}\frac{{\partial {w_i}}}{{\partial \xi }} + m\frac{{\partial \varphi }}{{\partial \xi }} + {\varepsilon ^2}m\frac{{\partial {Q_0}}}{{\partial \xi }}\frac{{\partial \varphi }}{{\partial \eta }} + m\frac{{\partial \varphi }}{{\partial \eta }} + {\varepsilon ^2}m\frac{{\partial {P_0}}}{{\partial \eta }}\frac{{\partial \varphi }}{{\partial \xi }} +  \\
m\frac{\sigma}{n_i}\frac{{\partial {n_i}}}{{\partial \xi }} + {\varepsilon ^2}m\frac{\sigma}{n_i}\frac{{\partial {Q_0}}}{{\partial \xi }}\frac{{\partial {n_i}}}{{\partial \eta }} + m\frac{\sigma}{n_i}\frac{{\partial {n_i}}}{{\partial \eta }} + {\varepsilon ^2}m\frac{\sigma}{n_i}\frac{{\partial {P_0}}}{{\partial \eta }}\frac{{\partial {n_i}}}{{\partial \xi }} +\\ \ldots  = 0, \\
\end{array}
\end{equation}
\begin{equation}\label{strain5}
\begin{array}{l}
{\varepsilon ^2}\left[ {\frac{{{\partial ^2}\varphi }}{{\partial {\xi ^2}}} + \frac{{{\partial ^2}\varphi }}{{\partial {\eta ^2}}}} \right] - \left[1-\alpha - {n_i} + \frac{3}{2}(1 + \alpha {\sigma _F})\varphi  + \frac{3}{8}(1 - \alpha {\sigma _F}^2){\varphi ^2}\right] +  \\
\ldots  = 0. \\
\end{array}
\end{equation}

\newpage

\textbf{FIGURE CAPTIONS}

\bigskip

Figure 1

\bigskip

(Color online) (a) The variation of soliton amplitude with respect to magnetic field direction for two different fixed values of relative positron concentrations, $\alpha$, showing the critical magnetic field angle, $\gamma_{cr}$ and a critical relative positron concentration. (b) Comparison of the variation of soliton amplitude with respect to magnetic field direction for a fixed value of relative positron concentrations, $\alpha$, for normal degeneracy (dashed-line) and ultra-relativistic degeneracy (solid-line) cases. (c) The critical fractional positron density, $\alpha_{cr}$, for the case of normal degeneracy. (d) Variation of IASW amplitude with respect to fractional positron density for ultra-relativistic degeneracy case (no such critical value exists for this case). The values of $u_{\xi,0}=u_{\eta,0}=0.1$ are used for all plots in this figure. The dash sizes in plots (a) and (c) are appropriately related to the values of varied parameter.

\bigskip

Figure 2

\bigskip

(Color online) The variation of soliton width with respect to magnetic field direction for different fixed values of different plasma parameters, namely, (a) magnetic field strength, (b) fractional positron concentration, (c) relative ion-temperatures, $\sigma$, for other fixed plasma parameters. (d) Comparison of the variation of soliton width with respect to magnetic field direction for fixed values of relative positron concentrations, $\alpha$, and magnetic field strength, $\overline\omega$, for normal degeneracy (dashed-line) and ultra-relativistic degeneracy (solid-line) cases. (e) Variation of soliton width with respect to the magnetic field strength, $\overline\omega$, for different relative positron concentrations and for fixed other plasma parameters. The values of $\varepsilon=0.1$ and $u_{\xi,0}=u_{\eta,0}=0.1$ are used for all plots in this figure. The dash sizes in plots (a), (b), (c) and (e) are appropriately related to the values of varied parameter.

\bigskip
\bigskip

Figure 3

\bigskip

(Color online) (a) The variation of collisional phase-shift with respect to magnetic field strength for fixed values of relative positron concentrations, $\alpha$, and fractional ion temperature, $\sigma$, for normal degeneracy (dashed-line) and ultra-relativistic degeneracy (dashed-line). (b) The dependence of collisional phase-shift on magnetic filed direction for non-relativistic degeneracy. (c) and (d), respectively show the dependencies of collisional phase-shift on relative positron concentrations, $\alpha$, and fractional ion temperature, $\sigma$ (normal degeneracy (dashed-line) and ultra-relativistic degeneracy (dashed-line)). The values of $u_{\xi,0}=u_{\eta,0}=0.1$ are used for all plots in this figure.

\bigskip

Figure 4

\bigskip

(Color online) Different regions in grey in $\alpha$-$\sigma$ plane showing where the IA solitons are bright for (a) normal degeneracy case, (b) ultra-relativistic degeneracy case and where the collision phase-shift is positive for (c) normal degeneracy case, (d) ultra-relativistic degeneracy case. The values of $\varepsilon=0.1$ and $u_{\xi,0}=u_{\eta,0}=0.1$ are used for all plots in this figure.

\bigskip

\newpage

\begin{figure}[ptb]\label{Figure1}
\includegraphics[scale=.6]{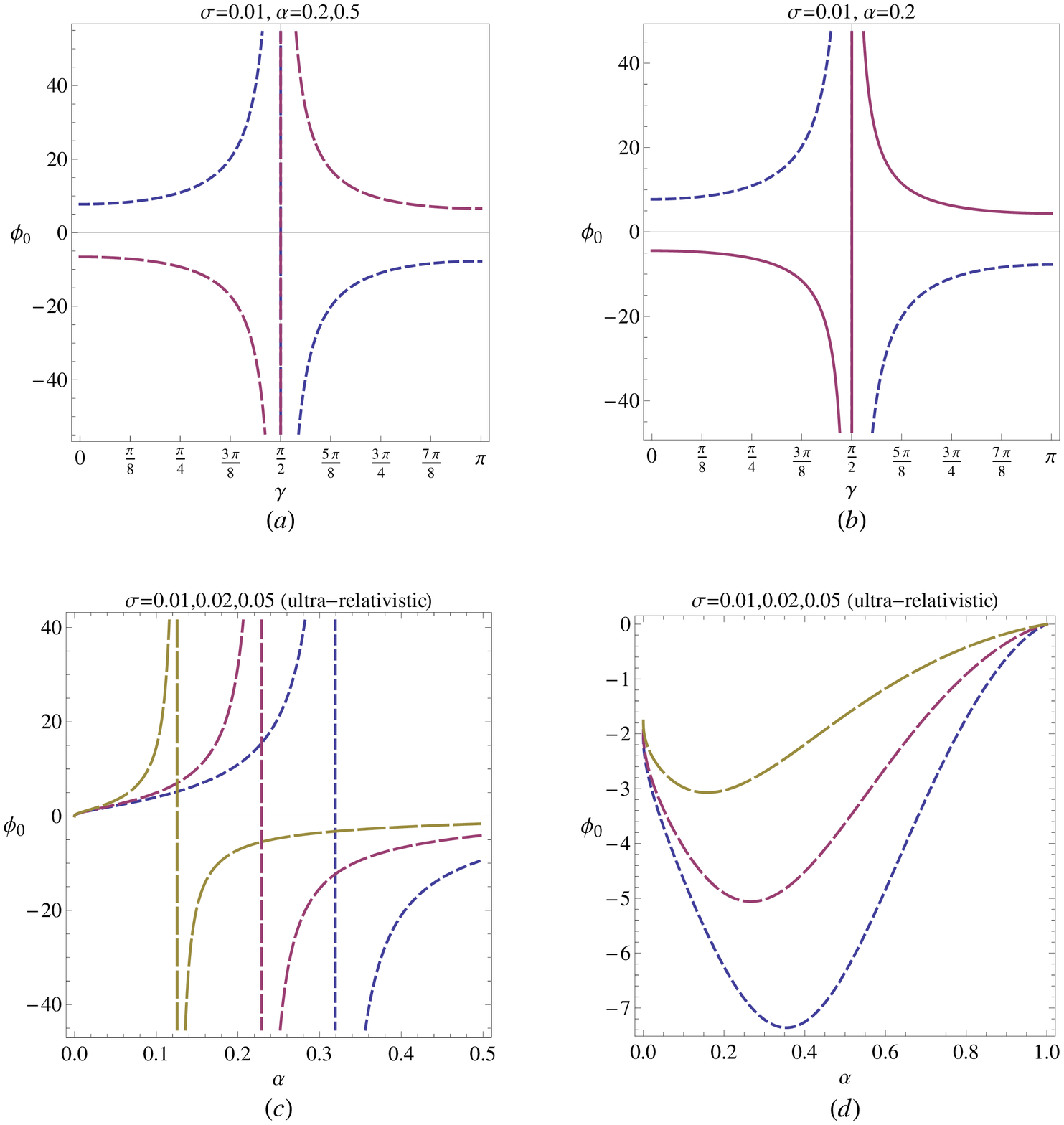}\caption{}
\end{figure}

\newpage

\begin{figure}[ptb]\label{Figure2}
\includegraphics[scale=.6]{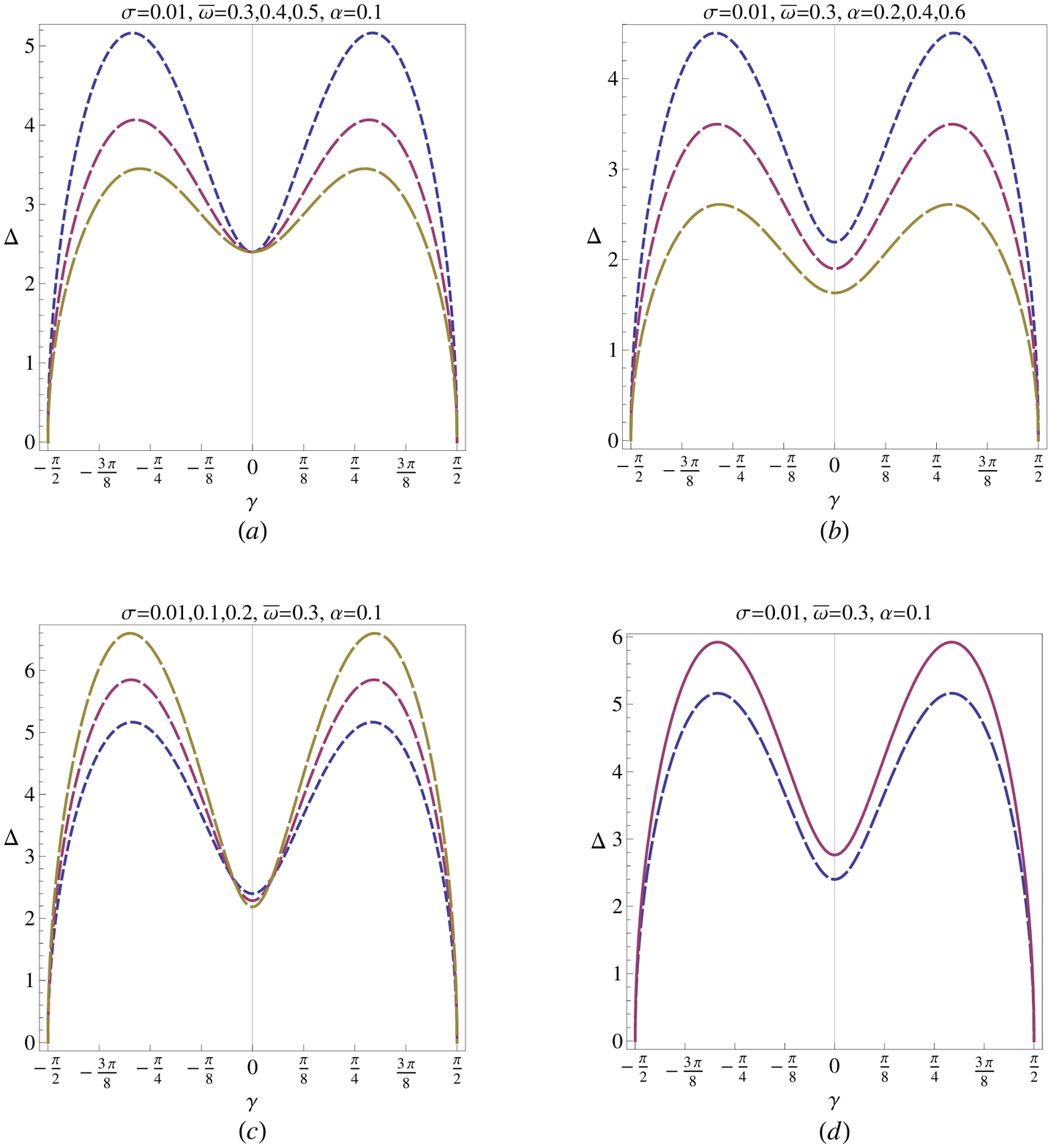}\caption{}
\end{figure}

\newpage

\newpage

\begin{figure}[ptb]\label{Figure3}
\includegraphics[scale=.6]{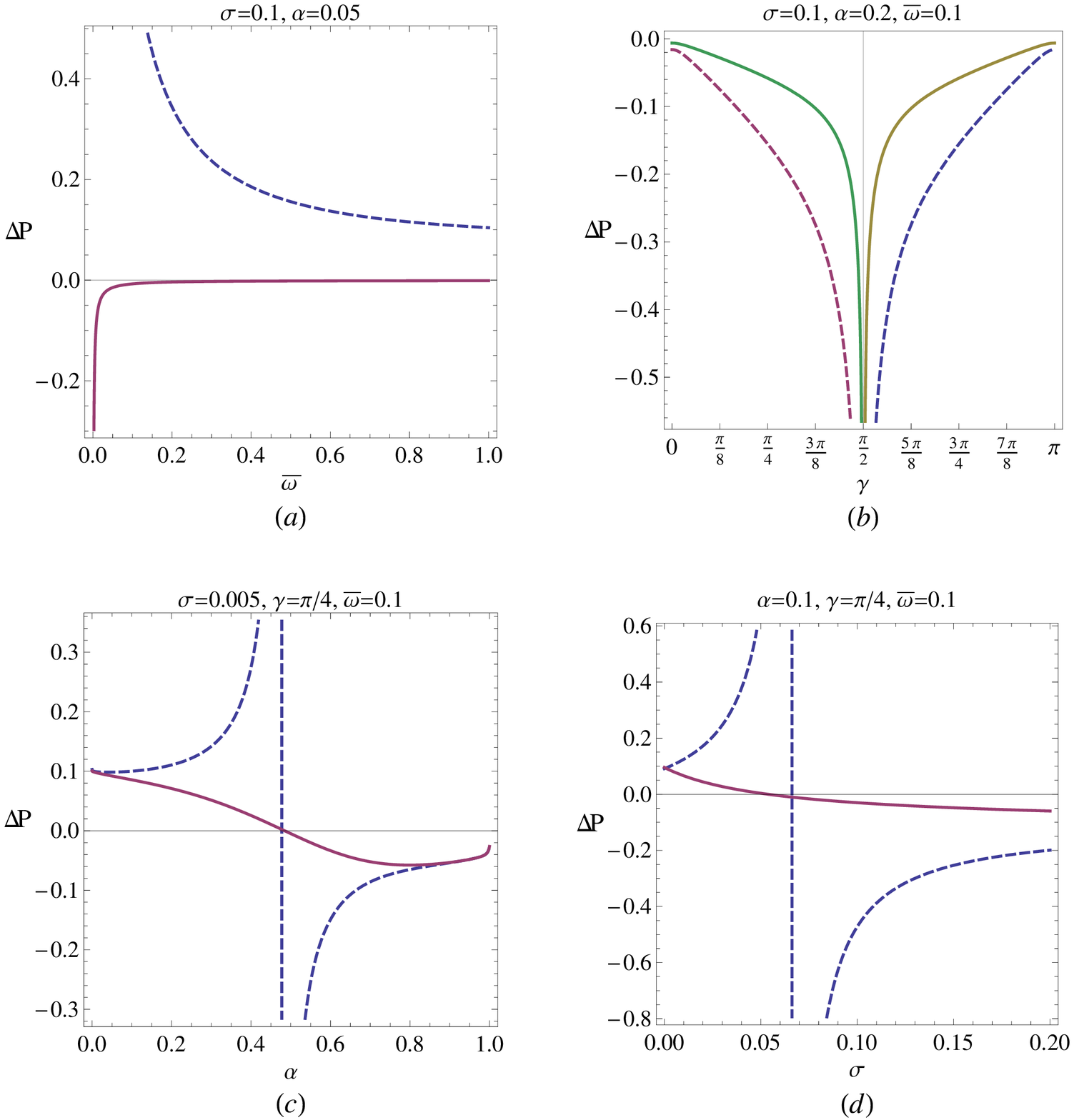}\caption{}
\end{figure}

\newpage

\begin{figure}[ptb]\label{Figure4}
\includegraphics[scale=.6]{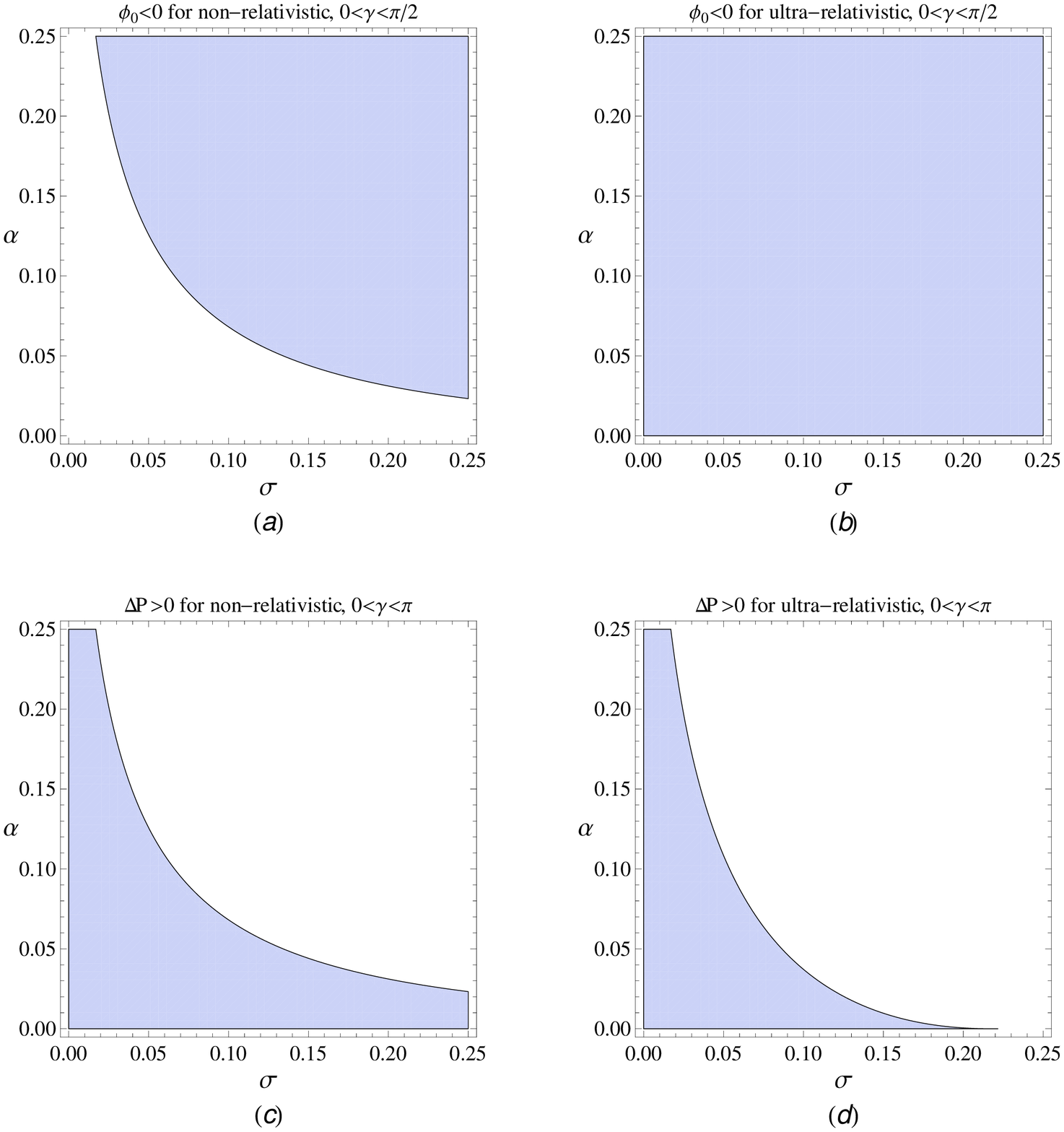}\caption{}
\end{figure}

\newpage

\end{document}